\documentclass{optica-article}

\journal{opticajournal} 

\articletype{Research Article}

\usepackage{lineno}
\usepackage{booktabs}
\begin{document}

\title{Alignment-Free Optical Combining with Scalar-Message Expectation Propagation Detection for Distributed Multi-Aperture Coherent Reception}

\author{Mingwen Zhu,\authormark{1} Yiqun Zhang,\authormark{1} Liyu Sun,\authormark{1} Ming Liu,\authormark{1} Changwei Wang,\authormark{1} and Jiayin Xue\authormark{1,*}}

\address{\authormark{1}Pengcheng Laboratory, No. 6001 Shahe West Road, Nanshan District, Shenzhen, Guangdong 518055, China
\\
}

\email{\authormark{*}xuejy@pcl.ac.cn}

\begin{abstract*} 
We propose an alignment-free distributed multi-aperture optical receiver with a single coherent front end, in which the differential delays between apertures are large. The branch signals are optically combined without inter-branch delay or phase alignment. Large differential delays produce separated tap clusters, forming a sparse ISI channel with long memory. Scalar-message expectation propagation exploits this sparsity to recover the transmitted symbols. Simulations show lower BER than equal-area single-aperture reception and no clear BER degradation over the tested delay spans. As the maximum differential delay increases from 100.5 to 200.5 symbol periods, the channel span nearly doubles, while the factor-graph edge count increases by only about 1.5\%. The trade-off is higher BER than ideal coherent combining schemes in exchange for fewer coherent front ends than digital combining and no inter-branch delay or phase alignment before optical combining.

\end{abstract*}

\section{Introduction}

Space laser communication has the advantages of large bandwidth, narrow beam, and immunity to electromagnetic interference, and has become an important candidate technology for high-speed satellite communication and integrated space--terrestrial networks \cite{Kaushal2017OpticalCommunication}. However, a coherent receiver couples the received optical field into a single-mode fiber (SMF) and is therefore sensitive to any distortion of the received wavefront. Atmospheric turbulence causes optical intensity scintillation and wavefront distortion. The wavefront distortion further results in fluctuations of the single-mode fiber coupling efficiency, which may lead to large received-power fluctuations and even deep fading \cite{Dikmelik2005FiberCoupling}. The use of multiple spatially separated receiving apertures can provide spatial diversity because the turbulence effects at different receiving positions are not fully correlated. Therefore, multi-aperture reception is an effective method to improve the reliability and availability of satellite-to-ground optical links \cite{Aghajanzadeh2010DiversityMultiplexing,Khalighi2009FadingReduction,Li2016PerformanceAnalysisFiberBased}.

For coherent optical reception, multi-aperture signals can generally be combined in the optical or digital domain \cite{Yang2017CombiningModule,Liu2023MultiApertureCoherentDigitalCombining,Ju2024RealTimeDemonstration}. In optical coherent combining, the delay, phase, and, when necessary, amplitude or polarization of each branch are adjusted before the common receiver, so that the optical fields from different apertures can be coherently combined \cite{Yang2017CombiningModule,Yang2017AllFiberCombining}. This architecture allows multiple apertures to share one coherent receiver after optical combining, but its performance depends on stable delay and phase relationships among different branches. However, for distributed telescopes installed on independent tracking mounts, the branch delays are affected by the spatial baseline, fiber length, and hardware path difference. The differential delay may also vary with the satellite line-of-sight direction, which makes large-range delay compensation and accurate phase stabilization difficult to implement.

On the other hand, digital coherent combining keeps independent coherent observations for different apertures and performs branch combining in the digital domain \cite{Geisler2016DigitalCoherentCombining,Geisler2018GroundReceiverArchitectures,Liu2023MultiApertureCoherentDigitalCombining,Ju2024RealTimeDemonstration}. The branch delay, phase, and gain differences can be compensated or weighted digitally, and diversity combining methods such as maximal-ratio combining (MRC) can be applied \cite{Ma2015PerformanceAnalysisSatelliteGround,Johst2024ExperimentalDemonstration}. However, the required coherent receivers and digital processing resources increase with the number of receiving apertures.

\begin{figure}[!h]
    \centering
    \includegraphics[width=8cm]{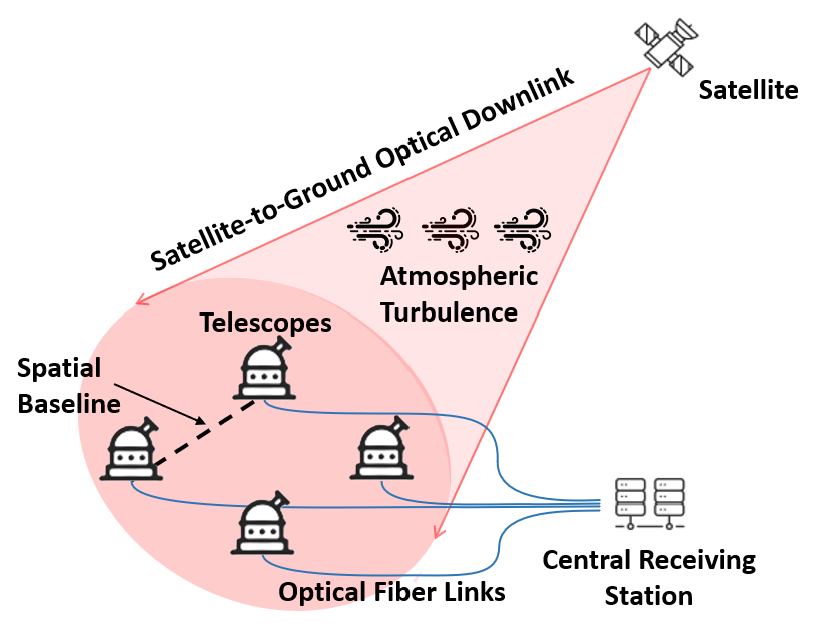}
    \caption{Distributed satellite-to-ground multi-aperture reception scenario with spatially separated telescopes connected to a central receiving station through optical fiber links.}
    \label{fig:scenario}
\end{figure}

Another approach is to combine the optical signals before a single coherent receiver and use digital equalization to handle intersymbol interference (ISI). Without relative delays, the branch signals overlap in time and may interfere destructively because of their random phases. To obtain diversity gain without adaptive optical phase and amplitude control, Belmonte introduced fixed branch delays of $0,T_s,\ldots,(L-1)T_s$~\cite{Belmonte2017DigitalEqualization}. These delays create artificial ISI, which is handled by digital equalization. Linear equalization, decision-feedback equalization, and maximum-likelihood sequence estimation (MLSE) were evaluated. In distributed satellite-to-ground reception, however, differential delays already arise from geometric path differences, unequal fiber lengths, and hardware paths. Fig.~\ref{fig:scenario} shows the considered reception scenario. Depending on the symbol rate, they can reach tens or hundreds of symbol periods and include fractional-symbol offsets. The resulting channel differs from the consecutive symbol-spaced channel considered by Belmonte. Applying conventional MLSE over the full channel memory would require a high computational cost. However, sufficiently separated branch delays produce local tap clusters, giving a channel with long memory but relatively few significant taps~\cite{Mietzner2006EqualizationSparseISI,Lee2004NonuniformlySpacedEqualizers}. This sparse structure can be used to reduce the detection complexity.

In this paper, we propose an alignment-free optical combining architecture with a single coherent front end for distributed satellite-to-ground multi-aperture reception with large differential delays. Optical signals from different telescopes are combined before the coherent receiver without inter-branch delay or phase alignment. The branch gains and delays are included in an equivalent sparse ISI channel with long memory. A scalar-message expectation propagation (EP)~\cite{Minka2001ExpectationPropagation} detector exploits this sparse structure to recover the transmitted symbols. Under different turbulence conditions and with perfect channel state information (CSI), EP is compared with linear minimum mean-square-error (LMMSE) detection and minimum mean-square-error decision-feedback equalization (MMSE-DFE). The receiver is also compared with equal-area single-aperture (SA) reception, ideal digital maximal-ratio combining (IMRC), and ideal optical coherent combining (IOCC). The results show a lower bit error rate (BER) than SA reception. EP generally performs better than LMMSE and MMSE-DFE at low and moderate signal-to-noise ratios (SNRs). We also examine how BER and computational cost change with the differential-delay span. The tested configurations show no clear BER degradation, and the EP edge count grows little once the local tap clusters are separated. The proposed receiver has higher BER than IMRC and IOCC, but uses fewer coherent front ends than IMRC and avoids the optical delay and phase alignment required by IOCC. It therefore provides a trade-off between BER performance and receiver hardware and control requirements.

\section{Distributed Multi-Aperture Reception Architectures}
\subsection{Differential Delay Characteristics of Distributed Multi-Aperture Reception}

In distributed multi-aperture reception, differential delays arise from geometric path differences, unequal fiber lengths, and hardware delays. Let $\mathbf{b}_{\ell}$ be the position of telescope $\ell$ relative to a reference telescope, and $\mathbf{u}(t)$ the unit vector pointing from the reference telescope toward the satellite. Under the far-field approximation, the geometric differential delay is
\begin{equation}
\Delta\tau_{\ell,\mathrm{geo}}(t)
\simeq
-\frac{\mathbf{b}_{\ell}^{\mathrm{T}}\mathbf{u}(t)}{c},
\end{equation}
where $c$ is the speed of light. Including the fiber and hardware contributions gives
\begin{equation}
\Delta\tau_{\ell}(t)
=
\Delta\tau_{\ell,\mathrm{geo}}(t)
+
\frac{n_g(L_{\ell}-L_{\mathrm{ref}})}{c}
+
\tau_{\ell,\mathrm{hw}}-\tau_{\mathrm{ref},\mathrm{hw}},
\end{equation}
where $n_g$ is the fiber group refractive index, $L_\ell$ is the branch fiber length, and $\tau_{\ell,\mathrm{hw}}$ is the hardware delay. The subscript $\mathrm{ref}$ denotes the reference branch. Depending on the path-length differences and symbol rate, the differential delays can span tens or hundreds of symbol periods and contain fractional-symbol offsets.

\subsection{Multi-Aperture Reception Architectures}
In conventional multi-aperture coherent reception, these branch differences are usually compensated either in the optical domain or in the digital domain.
\begin{figure}[!h]
\centering\includegraphics[width=13cm]{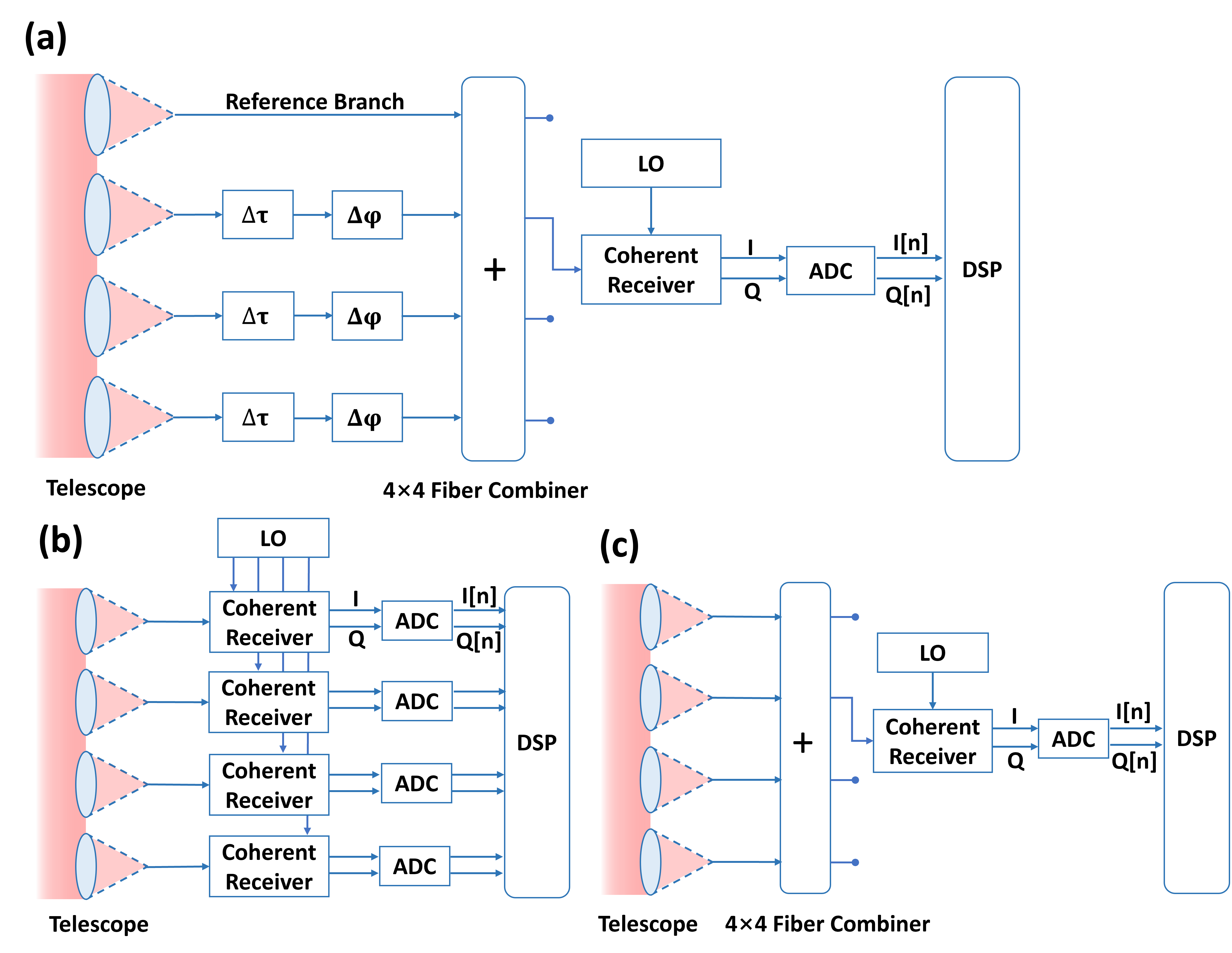}
\caption{Multi-aperture reception architectures. (a) Optical coherent combining; (b) Digital coherent combining; (c) Proposed alignment-free optical combining. $\Delta\tau$ and $\Delta\varphi$ denote the delay and phase compensation stages of each branch, respectively; LO, local oscillator.}
\label{fig:architectures}
\end{figure}

In optical coherent combining, the optical fields from different receiving apertures are combined before the coherent receiver, as shown in Fig.~\ref{fig:architectures}(a). The relative delay and carrier phase among different branches need to be compensated to ensure coherent addition of the optical fields. After optical combining, multiple apertures can share a common coherent receiver. This architecture reduces the number of coherent receiver chains, but requires accurate optical delay control and phase stabilization among different receiving branches.

In digital coherent combining, each receiving aperture is equipped with an independent coherent receiver, as shown in Fig.~\ref{fig:architectures}(b). The received signals from different branches are combined in the digital domain after coherent detection. The branch delay, phase, and amplitude differences can be compensated digitally. This architecture provides flexible signal processing capability, but the number of coherent receiver channels and the corresponding digital processing resources increase with the number of receiving apertures.

The proposed alignment-free optical combining architecture is shown in Fig.~\ref{fig:architectures}(c). The optical outputs from different receiving telescopes are directly combined and sent to a single coherent receiver without relative delay compensation and carrier-phase alignment. After coherent detection, the uncompensated branch gains and differential delays form a long-memory complex ISI channel in the digital domain, which is processed by receiver digital signal processing (DSP). The main hardware and processing requirements of the three multi-aperture reception architectures are summarized in Table~\ref{tab:architecture_comparison}.

\begin{table*}[!ht]
\centering
\small
\caption{Comparison of multi-aperture reception architectures.}
\label{tab:architecture_comparison}
\setlength{\tabcolsep}{4pt}
\begin{tabular}{lccc}
\hline
 & Optical CC & Digital CC & Proposed \\
\hline
Coherent front-ends
& 1 & $L$ & 1 \\

Optical-domain alignment
& Required & No & No \\

Independent branch observations
& No & Yes & No \\

Main processing
& Optical alignment & Digital alignment & Long-memory detection \\
\hline
\end{tabular}

\smallskip
{\small\raggedright \textit{Note:} CC, coherent combining.\par}
\end{table*}

\section{Signal and Equivalent Channel Model}

Consider a distributed multi-aperture coherent optical receiver with $L$ receiving branches. The optical signals from different telescopes are directly combined by a passive optical combining network without inter-branch delay compensation or carrier phase alignment. Only one output port is connected to a common coherent receiver front-end.
The branch gains and delays are assumed constant within each communication block. The equivalent complex baseband received signal is written as
\begin{equation}
r(t)
=
\sum_{\ell=1}^{L}
c_\ell \alpha_\ell s(t-\tau_\ell)
+n(t),
\end{equation}
where $s(t)$ is the transmitted complex baseband signal, $\alpha_\ell$ is the complex gain of the $\ell$-th branch, $\tau_\ell$ is its total propagation delay, and $c_\ell$ is the field transmission coefficient of combiner from the $\ell$-th input branch to the selected output port. The term $n(t)$ denotes the equivalent complex Gaussian noise of the coherent receiver.

The transmitted symbols $x[k]$ are drawn from a quadrature phase-shift keying (QPSK) constellation $\mathcal{A}=\{(\pm1\pm j)/\sqrt{2}\}$ with $\mathbb{E}\{|x[k]|^2\}=1$. After root-raised-cosine (RRC) pulse shaping, the transmitted signal is $s(t)=\sum_k x[k]p(t-kT_s)$, where $p(t)$ is the unit-energy RRC pulse and $T_s$ is the symbol period. For each receiving branch, the turbulence-induced complex gain is modeled as
\begin{equation}
\alpha_\ell
=
\sqrt{P_\ell}
\exp(j\theta_\ell),
\end{equation}
where $P_\ell$ is the optical power coupled into the SMF, and $\theta_\ell$ is the aperture piston phase caused by atmospheric turbulence.

For an ideal symmetric $L$-branch passive optical combining network with only one output port used, the field transmission coefficient of each branch is $c_\ell=1/\sqrt{L}$. 
The fixed phase introduced by the combining network can be included in $\alpha_\ell$, and thus $c_\ell$ is treated as a real-valued coefficient. For the four-aperture system considered in this paper, $c_\ell=1/2$. Since the branches are not aligned in delay and phase, the optical power from different branches cannot be fully concentrated at the selected output port. When only one output port is used and the time-averaged cross terms are small, the average power at the selected port can be approximated as $\mathbb{E}[P_{\mathrm{out}}]\approx\frac{1}{L}\sum_{\ell=1}^{L}P_\ell$. 

The discrete equivalent channel is derived based on the differential delays between branches. The branch with the earliest arrival time is selected as the delay reference, and its delay is denoted by $\tau_1$. The differential delay of the $\ell$-th branch is defined as $\Delta\tau_\ell=\tau_\ell-\tau_1$, where $\Delta\tau_1=0$. After receive matched filtering, the equivalent pulse response is $g(t)=p(t)*p^*(-t)$, and the symbol-rate equivalent channel tap is written as
\begin{equation}
h[q]
=
\sum_{\ell=1}^{L}
c_\ell\alpha_\ell
g(qT_s-\Delta\tau_\ell).
\end{equation}

\begin{figure}[!h]
    \centering
    \includegraphics[width=0.5\linewidth]{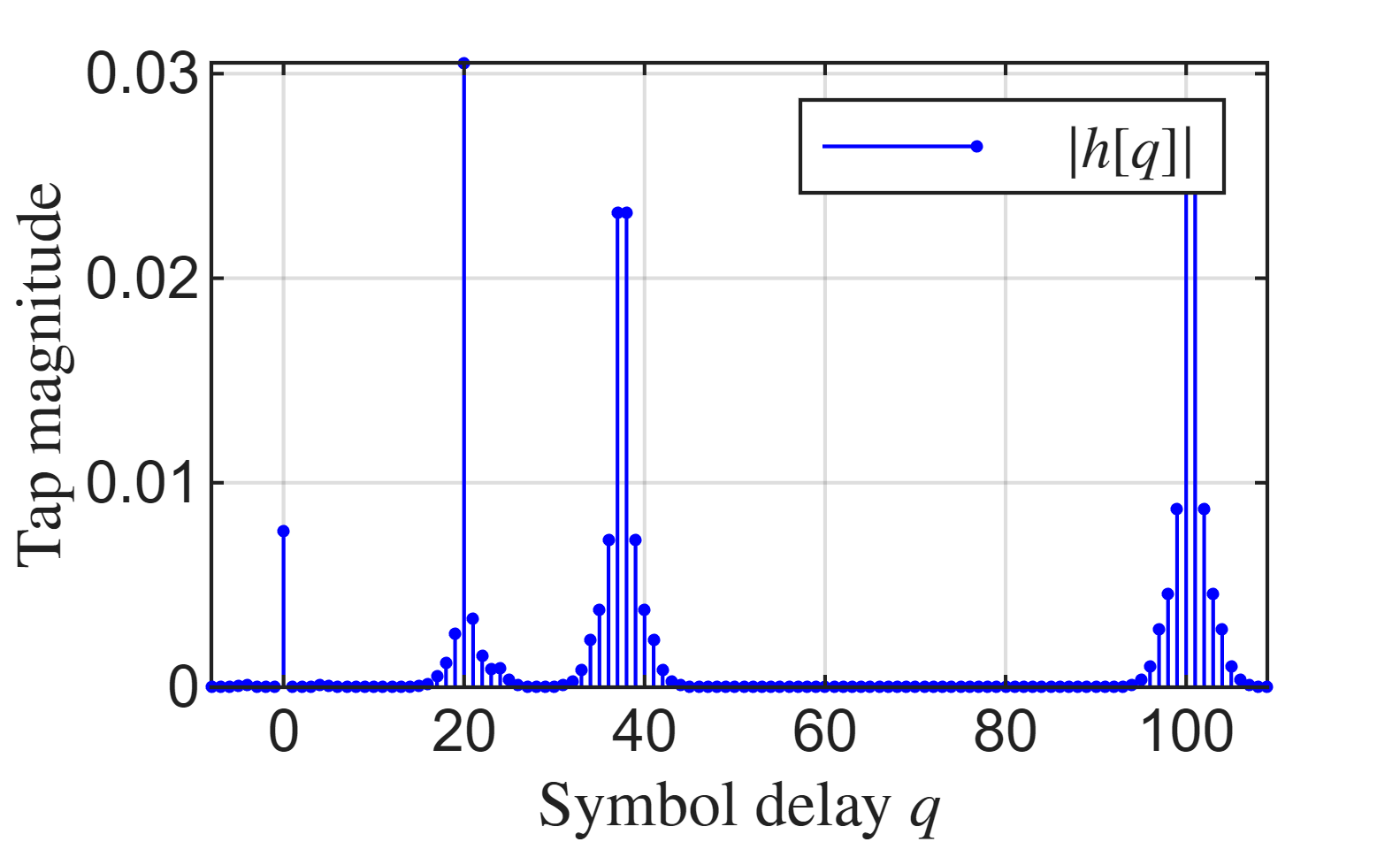}
    \caption{Equivalent discrete channel response $|h[q]|$ for one representative turbulence realization ($r_0 = 3.85$~cm). The differential delays are $[0,\ 20.1,\ 37.5,\ 100.5]T_s$. }
    \label{fig:channel_impulse_response}
\end{figure}

After sampling at $t=mT_s$, the discrete received signal is
\begin{equation}
z[m]
=
\sum_k h[m-k]x[k]
+w[m],
\end{equation}
where $w[m]$ is the complex Gaussian noise after matched filtering and sampling. When $\Delta\tau_\ell$ is close to an integer multiple of $T_s$, the energy of each branch is mainly concentrated around the corresponding tap position. For a fractional-symbol delay, the energy spreads over several neighboring taps due to pulse shaping and matched filtering, forming a local tap cluster. The differential delays mainly determine the separation between these clusters, while the cluster width is mainly determined by pulse shaping and fractional delay. Therefore, the equivalent channel has a large delay span with several separated groups of significant taps. A representative example is shown in Fig.~\ref{fig:channel_impulse_response}.

For a transmitted block of $N$ symbols, the linear-convolution model is written as
\begin{equation}
\mathbf{z}
=
\mathbf{H}\mathbf{x}
+
\mathbf{w},
\end{equation}
where $\mathbf{H}$ is the convolution matrix constructed from $h[q]$. If the tap-grid length is $K_h$, the received vector length is $M=N+K_h-1$. Since the local tap clusters are separated in time, $\mathbf{H}$ contains several separated nonzero diagonal bands. A larger differential-delay span increases $K_h$, but does not directly increase the number of nonzero elements in $\mathbf{H}$. This sparse structure is used in the scalar-message EP detector in the next section.

\section{Scalar-Message EP Detection}

EP has been applied to ISI equalization~\cite{Sun2015IterativeReceiverISI,Santos2017EPTurboEqualizer}. Here, scalar messages are used to exploit the sparse connections of the multi-cluster channel.

Based on the equivalent channel model, the factor graph contains $N$ variable nodes and $M$ factor nodes. Variable node $i$ represents the transmitted symbol $x_i$, and factor node $m$ corresponds to the received sample $z_m$. With $\mathcal{N}(m)$ denoting the set of variable nodes connected to $f_m$, the observation at $f_m$ is
\begin{equation}
z_m=\sum_{i\in\mathcal{N}(m)}h_{mi}x_i+w_m,
\end{equation}
where $h_{mi}$ is the $(m,i)$-th element of $\mathbf{H}$ and a connection exists when $h_{mi}\neq 0$.

The equivalent channel contains several separated tap clusters, so one transmitted symbol can affect received samples at distant time positions. The number of nonzero taps is still limited, and the factor graph therefore remains sparse. The total number of edges is $E=\operatorname{nnz}(\mathbf{H})=\sum_{m=1}^{M}|\mathcal{N}(m)|$, where $\operatorname{nnz}(\cdot)$ denotes the number of nonzero elements and $|\mathcal{N}(m)|$ the number of variable nodes connected to $f_m$.

The scalar-message EP detector updates messages only along these edges. In each iteration, the interference from the other symbols is approximated as a complex Gaussian variable~\cite{Wang2017DopedExpectationPropagation,Sun2015IterativeReceiverISI}. Let $\mu_{i\rightarrow m}$ and $v_{i\rightarrow m}$ denote the mean and variance of the message from variable node $i$ to factor node $m$. Factor node $m$ first accumulates
\begin{equation}
\bar{\mu}_m
=
\sum_{j\in\mathcal{N}(m)}
h_{mj}\mu_{j\rightarrow m},
\qquad
\bar{v}_m
=
\sum_{j\in\mathcal{N}(m)}
|h_{mj}|^2v_{j\rightarrow m}.
\end{equation}

then removes the contribution of $x_i$ and returns a message to variable node $i$. In natural parameters, this message reads
\begin{equation}
\lambda_{m\rightarrow i}
=
\frac{|h_{mi}|^2}{N_0+\bar{v}_m-|h_{mi}|^2v_{i\rightarrow m}},
\qquad
\eta_{m\rightarrow i}
=
\frac{
h_{mi}^{*}\!\left(z_m-\bar{\mu}_m+h_{mi}\mu_{i\rightarrow m}\right)
}{
N_0+\bar{v}_m-|h_{mi}|^2v_{i\rightarrow m}
}.
\end{equation}

Variable node $i$ sums the incoming factor messages, $\lambda_i^{\mathrm{ext}}=\sum_{m\in\mathcal{N}(i)}\lambda_{m\rightarrow i}$ and $\eta_i^{\mathrm{ext}}=\sum_{m\in\mathcal{N}(i)}\eta_{m\rightarrow i}$, and combines them with the constellation prior $p(a)$ to form the posterior
\begin{equation}
p_i(a)
\propto
p(a)
\exp\left[
-\lambda_i^{\mathrm{ext}}|a|^2
+
2\operatorname{Re}
\left\{
\eta_i^{\mathrm{ext}*}a
\right\}
\right],
\qquad a\in\mathcal{A}.
\end{equation}

The posterior mean and variance, $\hat{\mu}_i=\sum_{a\in\mathcal{A}}a\,p_i(a)$ and $\hat{v}_i=\sum_{a\in\mathcal{A}}|a-\hat{\mu}_i|^2p_i(a)$, follow from $p_i(a)$.

To avoid returning the information from factor node $m$ to itself, the target variable-to-factor message is obtained by message division:
\begin{equation}
\lambda_{i\rightarrow m}^{\mathrm{tar}}
=
\frac{1}{\hat{v}_i}
-
\lambda_{m\rightarrow i},
\qquad
\eta_{i\rightarrow m}^{\mathrm{tar}}
=
\frac{\hat{\mu}_i}{\hat{v}_i}
-
\eta_{m\rightarrow i}.
\end{equation}

A damping operation is applied to improve numerical stability:
\begin{equation}
\lambda_{i\rightarrow m}^{(t+1)}
=
(1-\beta)\lambda_{i\rightarrow m}^{(t)}
+
\beta\lambda_{i\rightarrow m}^{\mathrm{tar}},
\qquad
\eta_{i\rightarrow m}^{(t+1)}
=
(1-\beta)\eta_{i\rightarrow m}^{(t)}
+
\beta\eta_{i\rightarrow m}^{\mathrm{tar}},
\end{equation}
where $\beta$ is the damping factor. If an updated precision is non-positive or non-finite, the message from the previous iteration is retained. Iterations stop when the relative change in the posterior mean falls below a threshold or when the maximum number of iterations is reached. The detected symbol is obtained from the maximum posterior probability.

\section{Simulation Settings}

\subsection{Turbulence Conditions}

A quasi-static block-fading model is used. The branch gains and delays of different communication blocks are taken from the turbulence time series, that is, they are constant within one block and change between blocks. The Fried parameter $r_0$ is calculated from the Hufnagel--Valley (HV) turbulence profile~\cite{Valley1980IsoplanaticDegradation}. The wavefront phase is generated using random phase screens based on the modified von Karman spectrum~\cite{Sedmak2004FFTAtmosphericScreens}. The received power includes both SMF coupling~\cite{Dikmelik2005FiberCoupling} and Gamma--Gamma scintillation~\cite{AlHabash2001MathematicalModelIrradiance}. Pointing errors and platform vibrations are not considered.

The operating wavelength is $1550$~nm. Four subapertures with a diameter of $D_{\mathrm{sub}}=0.2$~m are placed at the corners of a square with a side length of $1$~m. The equal-area single-aperture reference has a diameter of $D_{\mathrm{single}}=0.4$~m. A common phase screen is translated across the four fixed subapertures at $12$~m/s. At each time step, the wavefront over each subaperture is sampled to calculate its piston phase and fiber coupling efficiency. The time series contains $N_t=4000$ states over $1.0$~s at a sampling rate of $4$~kHz. Weak, moderate, and strong turbulence conditions are generated by changing the $C_n^2$ profile. The corresponding values of $r_0$ and the aperture ratios are listed in Table~\ref{tab:turbulence_parameters}. 

\begin{table*}[!ht]
    \centering
    \small
    \caption{Parameters for the three turbulence conditions.}
    \label{tab:turbulence_parameters}
    \setlength{\tabcolsep}{4pt}
    \begin{tabular}{lccc}
        \hline
        Parameter & Weak & Moderate & Strong \\
        \hline
        $r_0$ (m) & 0.1545 & 0.0385 & 0.0166 \\
        $D_{\mathrm{sub}}/r_0$ & 1.29 & 5.19 & 12.05 \\
        $D_{\mathrm{single}}/r_0$ & 2.59 & 10.39 & 24.10 \\
        \hline
    \end{tabular}
\end{table*}

\begin{figure}[!h]
    \centering
    \includegraphics[width=\linewidth]{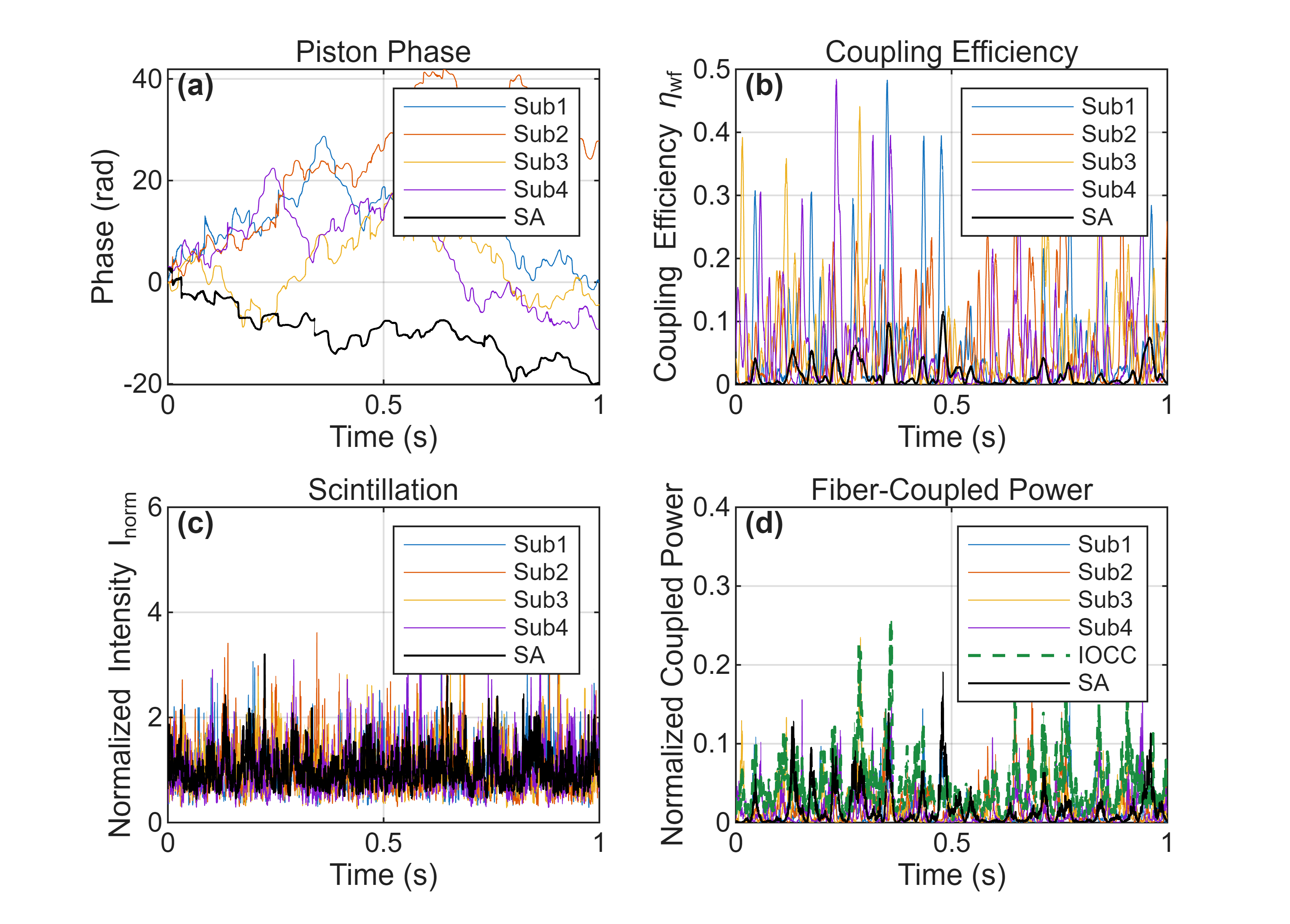}
    \caption{Turbulence characteristics under moderate turbulence: (a) piston phase; (b) single-mode fiber coupling efficiency; (c) intensity scintillation; and (d) fiber-coupled power. Sub1--Sub4 denote the four subapertures, SA denotes the equal-area single aperture, and IOCC denotes ideal optical coherent combining.}
    \label{fig:turbulence_characteristics}
\end{figure}

Fig.~\ref{fig:turbulence_characteristics} shows the piston phase, the fiber coupling efficiency, the intensity scintillation, and the fiber-coupled power of the subapertures and the SA under moderate turbulence. These quantities vary differently over time among the subapertures, which indicates that their turbulence disturbances are not fully correlated. This provides the basis for multi-aperture spatial diversity.

Fig.~\ref{fig:turbulence_characteristics}(b) shows the fiber coupling efficiency of the subapertures and the SA. The coupling efficiency of the SA is lower. This is because the $D_{\mathrm{sub}}/r_0$ of the subapertures is smaller, which reduces the effect of wavefront distortion on fiber coupling. Fig.~\ref{fig:turbulence_characteristics}(c) shows the intensity scintillation caused by turbulence. The intensity scintillation index is $0.1908$ for each subaperture, and it is $0.1054$ for the SA. The SA has weaker intensity scintillation, because its aperture averaging effect is stronger for a larger aperture. Fig.~\ref{fig:turbulence_characteristics}(d) shows the final received power of the subapertures and the SA. The SA has four times the area of one subaperture, so the input power is 1 for the SA and 0.25 for each subaperture. Thus, the SA and the four subapertures have the same total input power. This input power is defined before the intensity scintillation and the fiber coupling, and other factors that cause power attenuation are not considered here. The final received power is determined by both the intensity scintillation and the fiber coupling efficiency, and the two effects are combined as $P_{\mathrm{fiber}}(t)=P_{\mathrm{input}}(t)P_{\mathrm{scint}}(t)\eta_{\mathrm{SMF}}(t)$. The mean fiber-coupled powers of the four subapertures are $0.0140$, $0.0155$, $0.0156$, and $0.0150$. These values are close to the mean power of the SA, which is $0.0157$. Although the SA has a larger receiving area, its coupling efficiency is lower. Since the intensity scintillation is normalized, the mean received power is mainly determined by the coupling efficiency. Fig.~\ref{fig:turbulence_characteristics}(d) also shows the final received power of IOCC, which is given by the red curve. With ideal delay and phase alignment, equal-weight coherent combining gives a mean output power of $0.0452$, about $2.9$ times that of the SA. The total available power of the four apertures after the intensity scintillation and the fiber coupling loss is $0.0601$. Therefore, the average combining efficiency is $75.2\%$, which is below $100\%$ because the branch powers are unequal.

\subsection{Communication and Detection Parameters}

QPSK modulation is used with Gray mapping and four samples per symbol. The RRC filter has a roll-off factor of 0.20 and a span of eight symbol periods. Time is normalized by the symbol period $T_s$. One block of 512 symbols is simulated for each turbulence state. A total of $4.096\times10^6$ bits are simulated at each SNR point over 4000 states. The maximum number of EP iterations is 8, and the damping factor is 0.5.

The reference $E_b/N_0$ ranges from 0 to 30~dB in steps of 3~dB. We define the symbol energy of the SA as $E_s=1$. For QPSK, the energy per bit is $E_b=E_s/2$, so $E_s/N_0$ in dB is $3$~dB higher than $E_b/N_0$. Here, $N_0$ denotes the variance of the complex Gaussian noise on each received sample after matched filtering and symbol-rate sampling, where the in-phase and quadrature components each have a variance of $N_0/2$. For IMRC, each branch has independent complex Gaussian noise with variance $N_0$, and the branch signals and noise are combined using the MRC weights. The four branches require four coherent receivers, so the total receiver noise power is $4N_0$. For the single-front-end schemes, the received samples contain one complex Gaussian noise sequence with variance $N_0$, since all branches share one coherent receiver.

\section{Simulation Results and Discussion}

\subsection{BER Performance}

\subsubsection{BER under the Same Input Power Budget}
\begin{figure}[htbp]
    \centering
    \includegraphics[width=\linewidth]{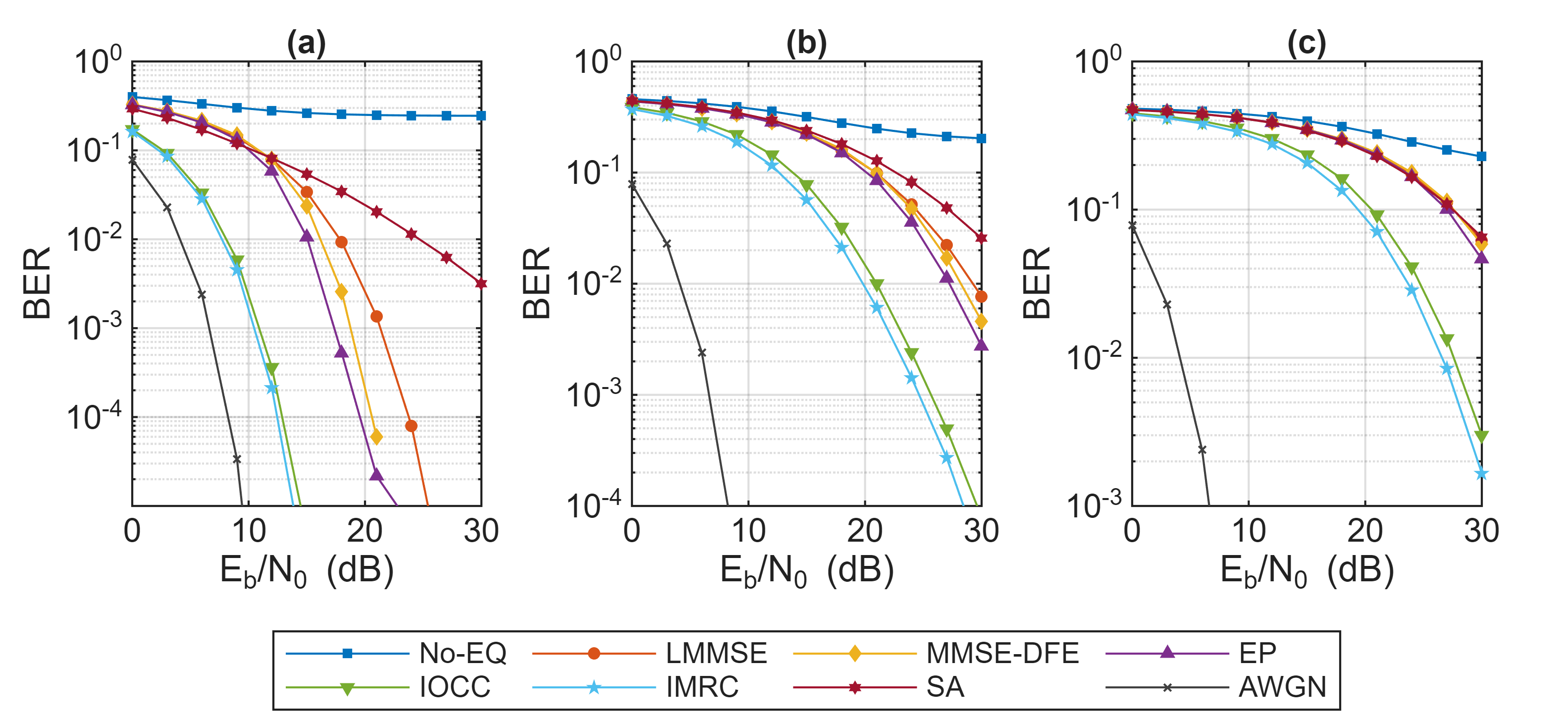}
    \caption{BER performance with perfect CSI under (a) weak turbulence with $r_0=15.45$~cm, (b) moderate turbulence with $r_0=3.85$~cm, and (c) strong turbulence with $r_0=1.66$~cm. IOCC denotes ideal optical coherent combining, IMRC denotes ideal digital maximal-ratio combining, and SA denotes the equal-area single aperture.}
    \label{fig:ber_turbulence}
\end{figure}

The BER is evaluated under weak, moderate, and strong turbulence with perfect CSI. The differential delays are fixed at $[0,\,5.1,\,9.5,\,25.5]T_s$. The EP detector is compared with reception without equalization (No-EQ), LMMSE, and MMSE-DFE using the same received samples, where No-EQ makes hard decisions directly, LMMSE uses block linear detection, and MMSE-DFE uses feedforward filtering and decision feedback~\cite{AlDhahir1995MMSEDecisionFeedback}. SA, IMRC, and IOCC are included as references, and the theoretical QPSK BER over a unit-gain AWGN channel is used as an additional reference. All schemes use the same total input optical power budget, turbulence states, and transmitted bits. Fig.~\ref{fig:ber_turbulence} compares their BER performance.

Under weak turbulence, IMRC and IOCC reach a BER of $10^{-3}$ at about $10.5$ and $10.9$~dB, respectively. The alignment-free receiver with EP reaches a BER of $10^{-3}$ at about $17.4$~dB, requiring $1.4$~dB less than MMSE-DFE and $3.9$~dB less than LMMSE. At $24$~dB, the EP BER decreases to about $5.9\times10^{-6}$, but residual errors around $10^{-6}$ remain at $27$--$30$~dB. In comparison, no errors are observed for MMSE-DFE at $24$~dB or for LMMSE at $30$~dB. EP therefore does not always perform better at high SNR. Compared with SA, EP gives a clear improvement. At $18$~dB, their BER values are $5.3\times10^{-4}$ and $3.4\times10^{-2}$, respectively. The SA BER remains at $3.2\times10^{-3}$ at $30$~dB, while No-EQ stays around $0.25$--$0.4$ over the tested SNR range.

Under moderate turbulence, IMRC and IOCC require about $24.6$ and $25.7$~dB to reach a BER of $10^{-3}$. At $30$~dB, their BER values are $3.4\times10^{-5}$ and $7.9\times10^{-5}$, respectively. EP achieves $2.7\times10^{-3}$, compared with $7.7\times10^{-3}$ for LMMSE and $4.6\times10^{-3}$ for MMSE-DFE. SA and No-EQ give $2.6\times10^{-2}$ and about $0.20$, respectively. None of the alignment-free detectors reaches $10^{-3}$ within the tested range. Under strong turbulence, all schemes show higher BER values. At $30$~dB, IMRC and IOCC achieve $1.7\times10^{-3}$ and $3.0\times10^{-3}$. EP gives $4.6\times10^{-2}$, while LMMSE, MMSE-DFE, and SA give $6.2\times10^{-2}$, $5.9\times10^{-2}$, and $6.5\times10^{-2}$, respectively. No-EQ remains at about $0.23$.

The alignment-free receiver with EP has a higher BER than IMRC and IOCC in all three turbulence cases. Under weak turbulence, EP requires about $6.9$ and $6.5$~dB more than IMRC and IOCC, respectively, to reach a BER of $10^{-3}$. Under moderate turbulence, both ideal references reach this BER within the tested range, while EP does not. Two effects contribute to this gap. IMRC retains independent branch observations and combines them digitally, whereas the alignment-free receiver uses only one output port of a passive combiner. Compared with IOCC, the absence of delay and phase alignment also reduces the power collected at this port. In addition, EP must detect symbols through the resulting ISI channel using approximate message updates and a limited number of iterations. 

\subsubsection{BER after Average Received Power Normalization}

This subsection examines why the alignment-free receiver with EP has a higher BER than IMRC and IOCC. To assess the contribution of the average received power difference, we normalize the average received signal powers to the same level under moderate turbulence and compare the remaining BER gap. The reference is $\bar{P}_{\mathrm{IMRC}}=\mathbb{E}\left[\sum_{\ell=1}^{4}P_\ell\right]$, the average total signal power received by the IMRC branches before digital combining. Let $\bar{P}_{\mathrm{EP}}$ and $\bar{P}_{\mathrm{IOCC}}$ be the output power of the proposed scheme and IOCC. The signal amplitude is multiplied by $\gamma_{\mathrm{EP}}=\sqrt{\bar{P}_{\mathrm{IMRC}}/\bar{P}_{\mathrm{EP}}}$ for EP and $\gamma_{\mathrm{IOCC}}=\sqrt{\bar{P}_{\mathrm{IMRC}}/\bar{P}_{\mathrm{IOCC}}}$ for IOCC. The channel is also scaled, but the noise is not changed. The scaling factor is fixed for all states, so the power fluctuation between states is kept. Under moderate turbulence, $\bar{P}_{\mathrm{EP}}$ is about $6.24$~dB below $\bar{P}_{\mathrm{IMRC}}$, and $\bar{P}_{\mathrm{IOCC}}$ is about $1.24$~dB below. The amplitude factors are $\gamma_{\mathrm{EP}}=2.0523$ and $\gamma_{\mathrm{IOCC}}=1.1529$.

\begin{figure}[!htbp]
    \centering
    \includegraphics[width=10cm]{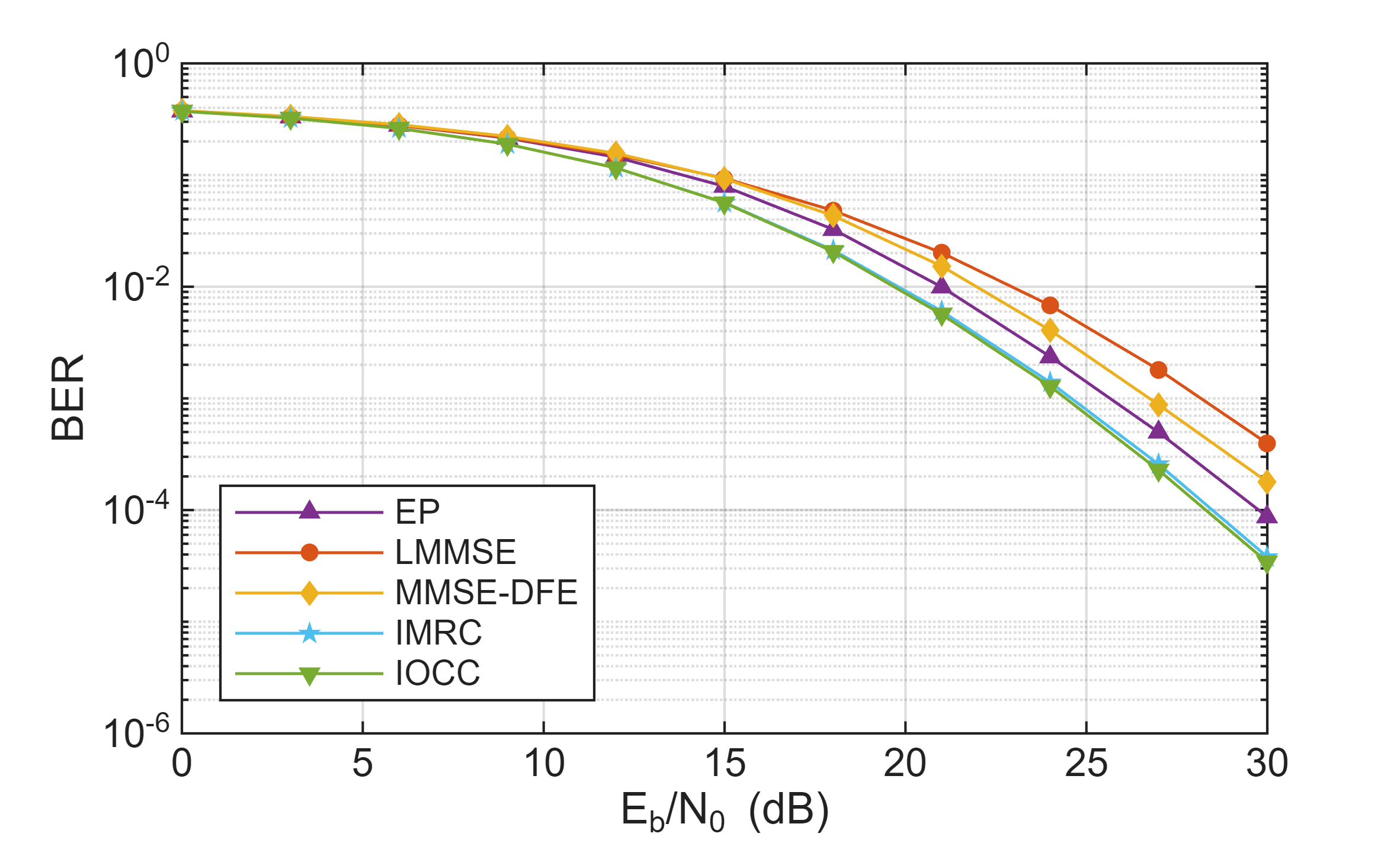}
    \caption{BER comparison under moderate turbulence after normalization to the same average effective received signal power.}
    \label{fig:ber_power_normalized}
\end{figure}

As shown in Fig.~\ref{fig:ber_power_normalized}, average-power normalization reduces the BER gap between EP and the two ideal references, while the IMRC and IOCC curves almost overlap. At $E_b/N_0=15$, $24$, and $30$~dB, the ratio of EP BER to IMRC BER decreases from about $3.9$, $25.3$, and $79.5$ before normalization to $1.40$, $1.70$, and $2.29$ after normalization, respectively. At the same average received power, EP gives BER results closer to those of IMRC and IOCC. These results indicate that the proposed architecture can provide aperture diversity gain, but the optical power loss from using only one coupler output port accounts for a large part of its BER gap relative to IMRC and IOCC.

\subsection{Differential-Delay Span and Complexity}

\begin{figure}[!h]
    \centering
    \includegraphics[width=10cm]{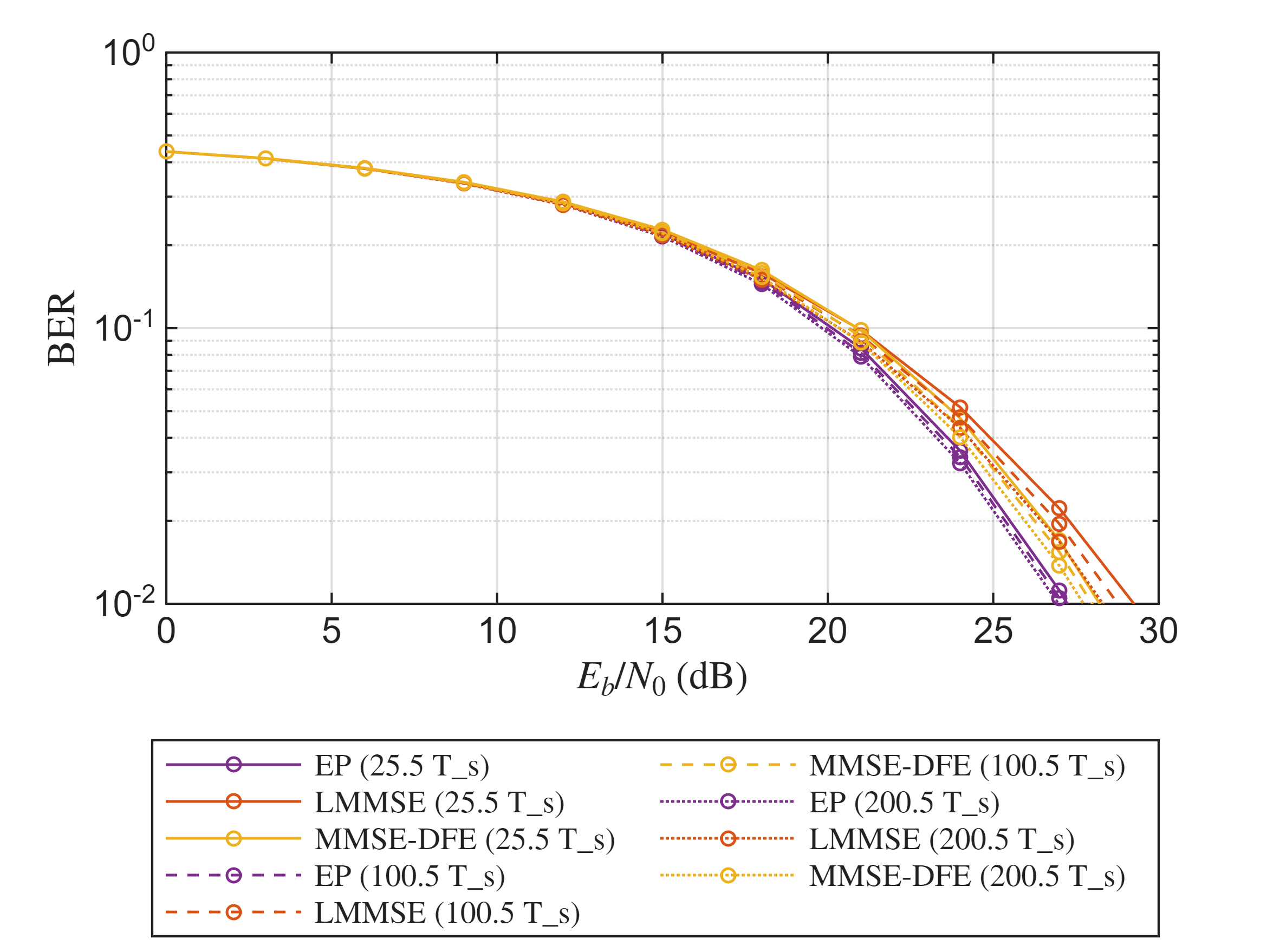}
    \caption{BER performance under moderate turbulence with perfect CSI for maximum differential delays of $25.5T_s$, $100.5T_s$, and $200.5T_s$.}
    \label{fig:delay_span_ber}
\end{figure}

Fig.~\ref{fig:delay_span_ber} compares EP, LMMSE, and MMSE-DFE under moderate turbulence with perfect CSI. The three configurations use the delay vectors $[0,\,5.1,\,9.5,\,25.5]T_s$, $[0,\,20.1,\,37.5,\,100.5]T_s$, and $[0,\,40.1,\,74.5,\,200.5]T_s$, that is, maximum differential delays of $25.5T_s$, $100.5T_s$, and $200.5T_s$. The EP curves remain close, while LMMSE and MMSE-DFE show slightly lower BERs at larger delay spans. These configurations mainly increase the separation between tap clusters, although some fractional-delay offsets also change. No clear BER degradation is observed for EP over the three tested configurations as the delay span increases.

We next examine the computational cost for the same three delay configurations. Let $I$ denote the average number of EP iterations, $E$ the number of effective factor-graph edges, and $N$ the number of data symbols in each block. Each factor node first calculates the total mean and variance, then subtracts the contribution of the target edge. The message updates require $O(E)$ operations per iteration. Each iteration also computes the QPSK posterior probabilities, mean, and variance for each symbol and checks convergence. These operations require $O(N)$ computations. The total complexity is therefore
\begin{equation}
C_{\mathrm{EP}}=O\!\left(I(E+N)\right).
\label{eq:ep_complexity}
\end{equation}
For the three differential-delay configurations, the average number of edges per symbol, $K_{\mathrm{eff}}=E/N$, equals 42, 69, and 70, respectively. Edge message updates dominate the computational cost, giving a leading complexity of $O(IE)$. Here, $K_{\mathrm{eff}}$ counts the effective connections per symbol rather than the tap-grid length $K_h$.

Table~\ref{tab:complexity_metrics} compares the three delay configurations. As the maximum delay increases from $25.5T_s$ to $200.5T_s$, $K_h$ increases from 43 to 218, while $E$ increases from 21\,504 to 35\,840. From $100.5T_s$ to $200.5T_s$, $K_h$ nearly doubles, but $E$ increases by only about $1.5\%$. Once the local tap clusters are separated, increasing their spacing adds few effective connections. At $E_b/N_0=12$~dB, the average iteration count is 8.00 for all three configurations, with almost all blocks reaching the iteration limit. At this SNR, $IE$ increases by a factor of about $1.67$ from the shortest to the longest delay span. 

\begin{table*}[!ht]
    \centering
    \small
    \caption{Computational-scale metrics for different differential-delay spans. At $E_b/N_0=12$~dB, the average EP iteration count is 8.00 for all three configurations, with almost all blocks reaching the iteration limit.}
    \label{tab:complexity_metrics}
    \setlength{\tabcolsep}{4pt}
    \begin{tabular}{cccccccc}
        \hline
        Maximum delay
        & $K_h$
        & $E$
        & $K_{\mathrm{eff}}$
        & $I$
        & $IE$
        & $\operatorname{nnz}(\mathbf{H}^{H}\mathbf{H})$
        & $\operatorname{nnz}(\mathbf{R})$ \\
        \hline
        $25.5T_s$
        & 43
        & 21\,504
        & 42
        & 8.00
        & $1.72\times10^5$
        & 40\,774
        & 20\,643 \\
        $100.5T_s$
        & 118
        & 35\,328
        & 69
        & 8.00
        & $2.83\times10^5$
        & 105\,724
        & 53\,118 \\
        $200.5T_s$
        & 218
        & 35\,840
        & 70
        & 8.00
        & $2.87\times10^5$
        & 153\,658
        & 87\,436 \\
        \hline
    \end{tabular}
\end{table*}

The LMMSE and MMSE-DFE implementations used here both construct $\mathbf{A}=\mathbf{H}^{H}\mathbf{H}+N_0\mathbf{I}$ and perform sparse Cholesky factorization. Here, $\mathbf{R}$ denotes the Cholesky factor of $\mathbf{A}$. LMMSE uses triangular solves after factorization, while MMSE-DFE also applies sequential decision feedback. Their computational costs depend on the nonzero pattern of the Gram matrix and the additional nonzero entries created during Cholesky factorization. From the shortest to the longest delay span, $\operatorname{nnz}(\mathbf{H}^{H}\mathbf{H})$ increases from 40\,774 to 153\,658, a factor of about $3.77$. The number of nonzero elements in $\mathbf{R}$ increases from 20\,643 to 87\,436, a factor of about $4.24$. Although these matrices remain $N\times N$, their nonzero counts grow with the delay span. For the tested configurations, the EP edge count grows more slowly with delay span than the nonzero counts of the Gram matrix and its Cholesky factor.

\section{Conclusion}

We proposed an alignment-free optical combining architecture for distributed satellite-to-ground multi-aperture reception. The aperture signals are combined before a single coherent front end without inter-branch delay or phase alignment. Instead of compensating the branch differences before optical combining, the receiver includes them in an equivalent multi-cluster ISI channel and recovers the symbols through digital detection. Scalar-message EP is used here to handle this channel by exploiting its sparse structure.

With perfect CSI, the proposed receiver achieves a lower BER than equal-area single-aperture reception under the three tested turbulence conditions, showing its ability to provide diversity gain. No clear BER degradation is observed as the maximum differential delay increases from $25.5T_s$ to $200.5T_s$. Once the local tap clusters are separated, a larger delay span adds few effective channel connections, so the message-update cost grows little. These results support the use of this architecture for multi-aperture reception with large differential delays. Under the same input power budget, its BER remains higher than those of IMRC and IOCC. After average received power normalization, however, its BER approaches those of both ideal references. This indicates that the lower average received power accounts for a large part of the BER gap under the same input power budget. The architecture therefore trades BER performance for fewer coherent front ends than digital combining and no inter-branch optical delay or phase alignment before optical combining.
\section*{Funding}
Mobile Information Networks-National Science and Technology Major Project under Grant no. 2025ZD1302900.

\section*{Disclosures}
The authors declare no conflicts of interest.

\section*{Data Availability}
Data underlying the results presented in this paper are not publicly available at this time but may be obtained from the authors upon reasonable request.

\bibliography{sample}

\end{document}